# A versatile method for nano-fabrication on diamond film: flexible diamond metasurfaces as a demonstration

*Yicheng Wang[#], Jixiang Jing[#], Yumeng Luo, Linjie Ma, Zhongqiang Wang, Qi Wang\*, Kwai Hei Li\*, Zhiqin Chu\**

Yicheng Wang, Jixiang Jing, Linjie Ma, Zhiqin Chu

Department of Electrical and Electronic Engineering, The University of Hong Kong, Pokfulam, Hong Kong.

E-mail: zqchu@eee.hku.hk

Yumeng Luo, Kwai Hei Li

School of Microelectronics, Southern University of Science and Technology, Shenzhen 518055, China.

E-mail: khli@sustech.edu.cn

Zhongqiang Wang, Qi Wang

Dongguan Institute of Opto-Electronics, Peking University, Dongguan 523808, China.

E-mail: wangq@pku-ioe.cn

# These authors have contributed equally to this work.

* Corresponding authors

**Abstract:**

Diamond exhibits superb performance across a wide range of applications due to its enormous outstanding properties in electronic, photonic and quantum fields. Yet heterogeneous integration of diamond for on-chip functionalities, like 2D materials, remains challenging due to the hard acquisition of scalable, transferable and ultrathin diamond samples. Recently, the edge-exposed exfoliation has been demonstrated as an effective way to produce wafer-scale, freestanding and ultrathin diamond films. However, the incompatibility of the newly developed diamond film with conventional nano-fabrication methods makes it difficult to fabricate diamond film into practical devices. Herein, we demonstrate the mask-transferring by sugar as a versatile method for pattern-definition on diamond films, which shows excellent geometrical resolution and accuracy comparing to conventional approaches. Additionally, based on this method, the flexible all-diamond metasurfaces functioning as structural colors have been achieved, which indicates its huge potential for fabricating more diamond-related devices.

## 1. Introduction

Diamond is widely proposed for future electronic and photonic technologies due to its superlative material properties including large bandgap, high carrier mobility, thermal conductivity,[1,2] and ultrawide optically-transparent window from infrared to ultraviolet. However, heteroepitaxial growth of diamond on arbitrary substrates remains difficult, hindering the integration and evolution of diamond-based technologies. To build diamond heterogeneous platforms for expanding on-chip functionalities, including electronics (e.g., transistors,[3-6] P-N junctions[7,8]), photonics (e.g., waveguides,[9-11] resonators,[12-14] metasurfaces,[15-18] metalens[19-21]), and quantum sensors (e.g., thermometers,[22-25] magnetometers,[26-28]), the integrable and ultrathin diamond is urgently needed. Despite extensive efforts have been made over the past decades, producing large quantities of desired ultrathin and transferable diamond film for widespread use still remains challenging.

Currently, the edge-exposed exfoliation using sticky tape has been demonstrated as a simple, scalable, and reliable method for massively producing ultrathin and transferable diamond films.[29] However, as newly developed three-dimensional (3D) film, it is still difficult for the post-manipulation of ultrathin diamond (e.g., on-surface nanofabrication) referring the way of two-dimensional (2D) materials and bulk materials via conventional methods (e.g., lithography and etching). The incompatibility of diamond film with conventional nanofabrication process mainly attribute to its special properties, including i) low surface electrical conductivity, which suffers from proximity effect and charge accumulation, easily inducing the aggregation of electrons when conducting electron-beam-lithography (EBL), diminishing the accuracy of defined pattern; ii) fragility that leaves shattering and cracking due to the stress during directly spin-coating on diamond film, especially for films with thickness thinner than one micron, undermining its integrity; and iii) surface geometric fluctuation when attached to curved and flexible surfaces, which could easily leads to uneven distributions of spin-coated resists and simultaneously bring difficulties in sample focusing when implementing photo-/ e-beam lithography. Despite those strategies have been proposed to replace conventional wafer-fabrication processes including transfer printing,[30-32] metal or conductive polymer coating[33,34] for anti-charge accumulation, self-assembly nano spheres,[35,36] nano-imprint lithography.[37-39] However, those methods still encounter limitations such as uneven distribution of conductive polymer, chemical contaminations to diamond and substrate, and challenges in the removal of deposited metals without damaging diamond sample and substrate.

In this study, we demonstrate mask-transferring by sugar as a versatile method to help conduct high-precision, large-scale, repeatable nano-fabrication on our developed diamond films. Particularly, utilizing sugar as the transfer medium, we can successfully transfer the pre-defined patterns to diamond films as masks, and conduct diamond etching subsequently. This approach is straightforward, non-polluting, and highly accurate, unlocking the vast possibility of diamond film applications. As a-proof-of photonic application, the structural color based on all-diamond metasurfaces on flexible polyethylene-glycol-terephthalate (PET) substrate is showcased. Related characterizations indicate the diamond-based structural colors exhibit higher reflectance, better color saturation and wider color gamut than previous works,[16,40-42,44] which is also stable and robust under different extent of flexible deformations, making it possible for high-quality[45] and stable display[46,47] applications. This developed sugar-transfer method, opens a new way of high-precision nanofabrication on transferable diamond films, and will undoubtedly boost the development of diamond in diverse fields.

## 2. Results and discussion

### 2.1 A versatile method for nano-fabrication on diamond film

The diamond film with ~ 600 nm thickness was firstly grown by microwave plasma chemical vapor deposition (MPCVD) method on silicon substrate (see Methods). When attaching the sticky tape on the surface of diamond and stretching it at a moderate speed, the diamond film could be gradually peeled off from silicon substrate,[29] as shown in Figure. 1(a) and Figure. S1. Photographs of diamond-on-silicon assembly and exfoliated diamond film are shown in Figure 1(b) and 1(c), respectively, where 2-inch size was achieved. High-magnification scanning electron microscope (SEM) images (Figure. 1(d, e)) and atomic force microscope (AFM) images (Figure. 1(f, g)) present the surface morphology of two sides of exfoliated diamond films, and corresponding surface amplitude along the white dashed lines in Figure.

1(f), g are presented in Figure. 1(h) and 1(i). Compared with the rough as-grown (top) surface (Ra = 44.58 nm), the buried (bottom) surface exhibits extremely lower roughness (Ra: ~ 1nm) that is seldomly mentioned by others. This discovered ultra-flat surface without experiencing any post-flattening process (e.g., mechanical or chemical polishing) makes it accessible to high-precision nanofabrication, overcoming the bottleneck of the rough as-grown surface for widespread use in a long term (Figure. S2).

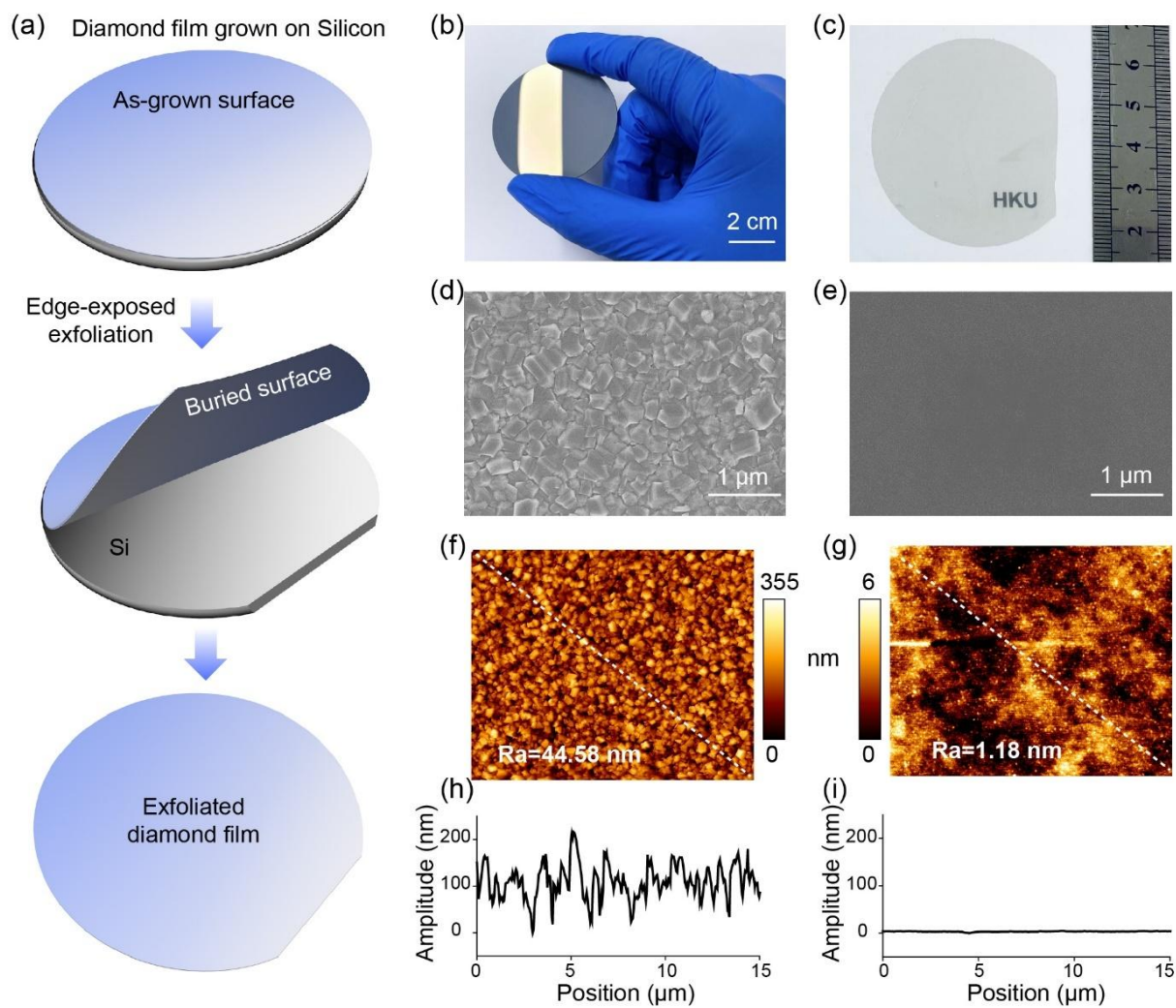

**Figure 1**. (a) Schematic diagram of exfoliating diamond film from silicon wafer. (b-c) Photograph of 2-inch diamond-on-silicon wafer (b) and exfoliated diamond film (c), respectively. (d-e) SEM images of as-grown surface (d) and buried surface (e) of diamond film. (f-g) AFM images of as-grown surface (f) and buried surface (g) of diamond film. (h-i) Corresponding surface amplitude along the white dashed lines in (f) and (g), respectively.

After acquiring the scalable, ultra-flat diamond film, it still remains difficult to conduct high-precision nanofabrication on its surface when following conventional fabrication methods including photolithography and E-beam lithography (Details see Supporting Information Section 3). To unleash the limitation of conventional wafer-fabrication process, we herein introduce a versatile method to conduct nanofabrication on diamond film by sugar transferring, which can achieve large-area, high-precision, and good repeatability. The schematic illustration of our proposed method is shown in Figure. 2, which can be generally classified into three steps: (i): Mask preparation: Utilizing E-beam lithography to acquire designed mask on indium-tin-oxide (ITO) film substrate (see method); (ii) Sugar transfer: to effectively transfer the fabricated pattern to diamond surface, a mixture of corn syrup, cane sugar and deionized water was prepared.[30] By dropping sufficient sugar solution (5-10 μL) on top of fabricated template pattern on ITO film, and then putting them inside of a heating oven at 70 °C for several hours, the sugar solution would be solidified. Notably, excessively low and high temperature could induce overlong solidifying time and residual bubbles, respectively, damaging the template pattern. Next, by slightly bending the PET film, the solid sugar would be easily detached from the film where the residual stress on dried sugar may assist the separation [30] without needing any sacrificial layer; (iii) Diamond etching: By placing the dried sugar on diamond film and gently pressing it with tweezer, diamond film would be conformed to the flat surface of sugar. The sugar was subsequently dissolved away with deionized water, leaving designed masks on diamond film. Inductive coupled plasma (ICP) was employed for diamond etching where the flow rate of oxygen gas, RF voltage, bias voltage, and cavity pressure was set to 80 sccm, 200 W, 60 W, and 10 mTorr, respectively. Finally, the fabricated diamond pattern would be obtained after removing the masks by KOH solution. Utilizing predefined mask on conductive ITO film, difficulties of direct exposure on diamond film (e.g., diamond film's fragility, uneven distribution of resist, and problems with focusing) can be avoided, and the entire transfer procedure is chemicals-free to diamond film and the substrate, which avoids contaminations from damaging the device and substrate.

WILEY-VCH

(i) Mask preparation

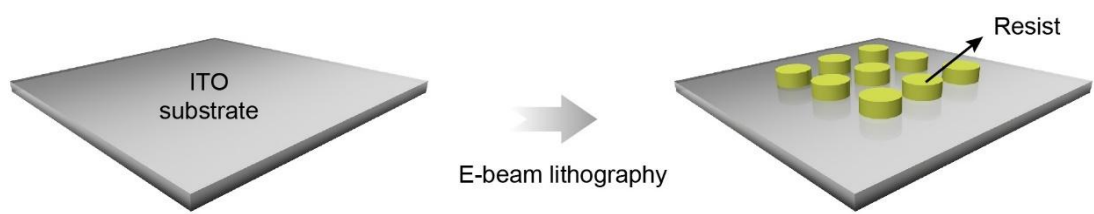

(ii) Sugar-transfer

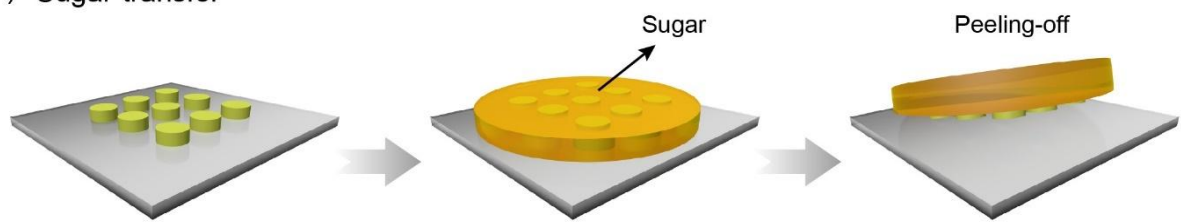

(iii) Diamond etching

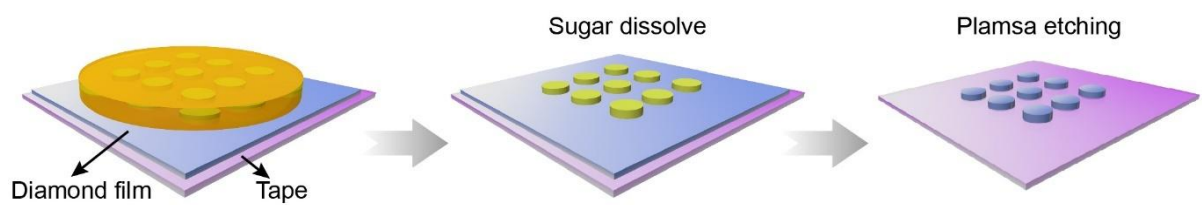

**Figure 2**. Schematic illustration of nano-fabrication on diamond film using our proposed sugar-transfer method.

## 2.2 Comparison between our proposed method and conventional method

Figure 3(a) shows the schematic illustration comparing mask preparation procedure using our proposed method and conventional method, where our proposed method utilized aforementioned sugar-transfer technique, and conventional method involved spin-coating of resist and E-beam lithography directly on diamond film. Figure 3(b-g) present the SEM images of fabricated masks on buried surface of diamond film utilizing our method (b-d) and conventional lithography method (e-f), respectively. Particularly, the circular pillar arrays by sugar-transferring (Figure. 3(b)) possess more regular morphology than that obtained by direct lithography (Figure. 3(e)). In addition, the fabricated nano-gratings also demonstrate the superiority of our proposed sugar-transferring method. Different magnification SEM images (Figure. 3(c, d)) indicate the well-defined morphology (e.g., accurate parameter as designed, and clear boundaries) of nano-gratings by sugar transferring. In contrast, nano-gratings obtained by conventional direct lithography (Figure. 3(f, g)) exhibit geometrical distortion that deteriorates from the edge of the pattern toward the center, which mainly attributes to charge accumulation and proximity effect during EBL exposure.

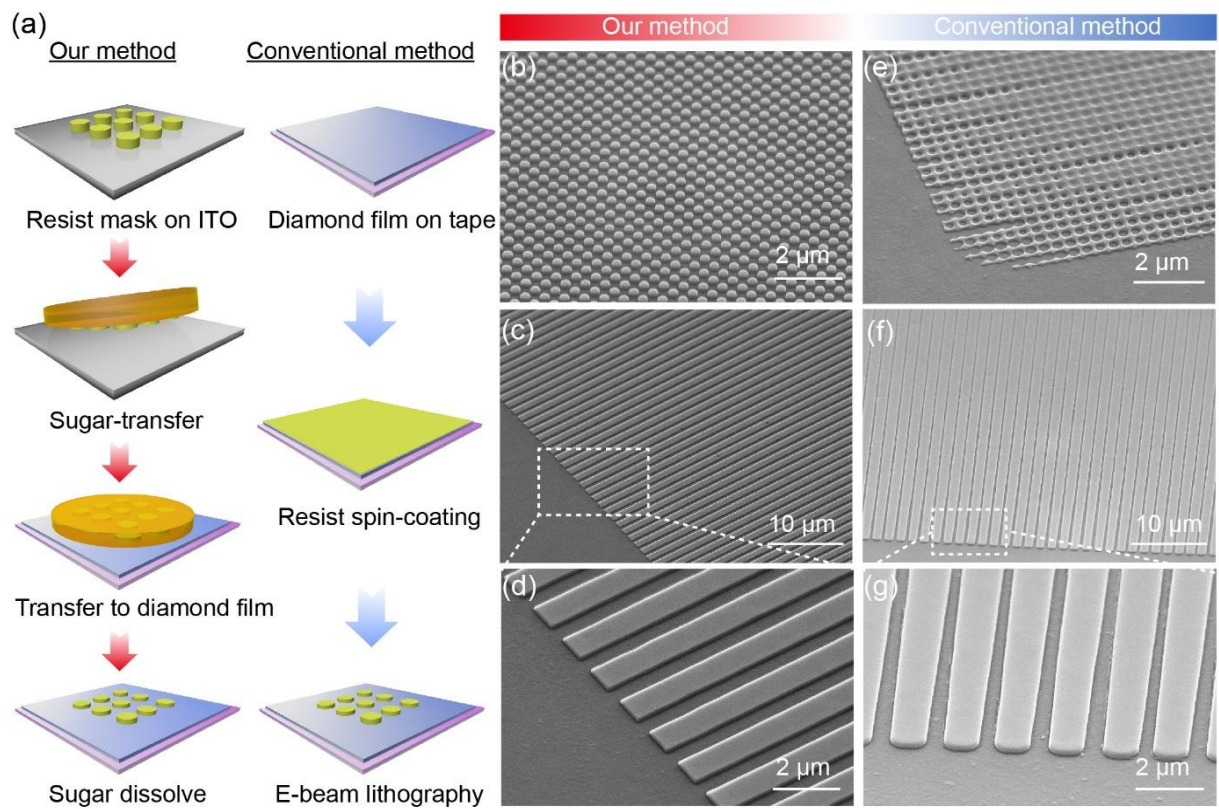

**Figure 3.** (a) Schematic illustration of our sugar-transferring method and conventional method for fabricating mask pattern on diamond film. (b, e) SEM images of circular array masks with the period of 400 nm made on diamond film's buried surface using our method (b) and conventional method (e), respectively. (c, f) SEM images of grating masks made on diamond film's buried surface using our method (c) and conventional method (f), respectively. (d, g) Zoomed in images of (c) and (f).

## 3. Flexible structural colors based on all-diamond metasurfaces

After demonstrating the superiority of our sugar-transferring method, the performance of fabricated diamond-based device should be further demonstrated. Despite the superior characteristics of diamond in the field of photonics and optics, to the best of our knowledge, diamond-based metasurfaces have been seldomly reported up to now. Therefore, the structural colors based on Mie's resonance effect is showcased for all-diamond flexible metasurfaces. Compared with traditional metallic structural colors who suffers from high absorption at blue region, and dielectric (e.g., Si, TiO$_2$) structural colors that are prone to corrosion, diamond material emerges as an outstanding candidate of photonic devices due to its high refractive index, negligible extinction coefficient,[29,44] and exceptional chemical and mechanical durability.

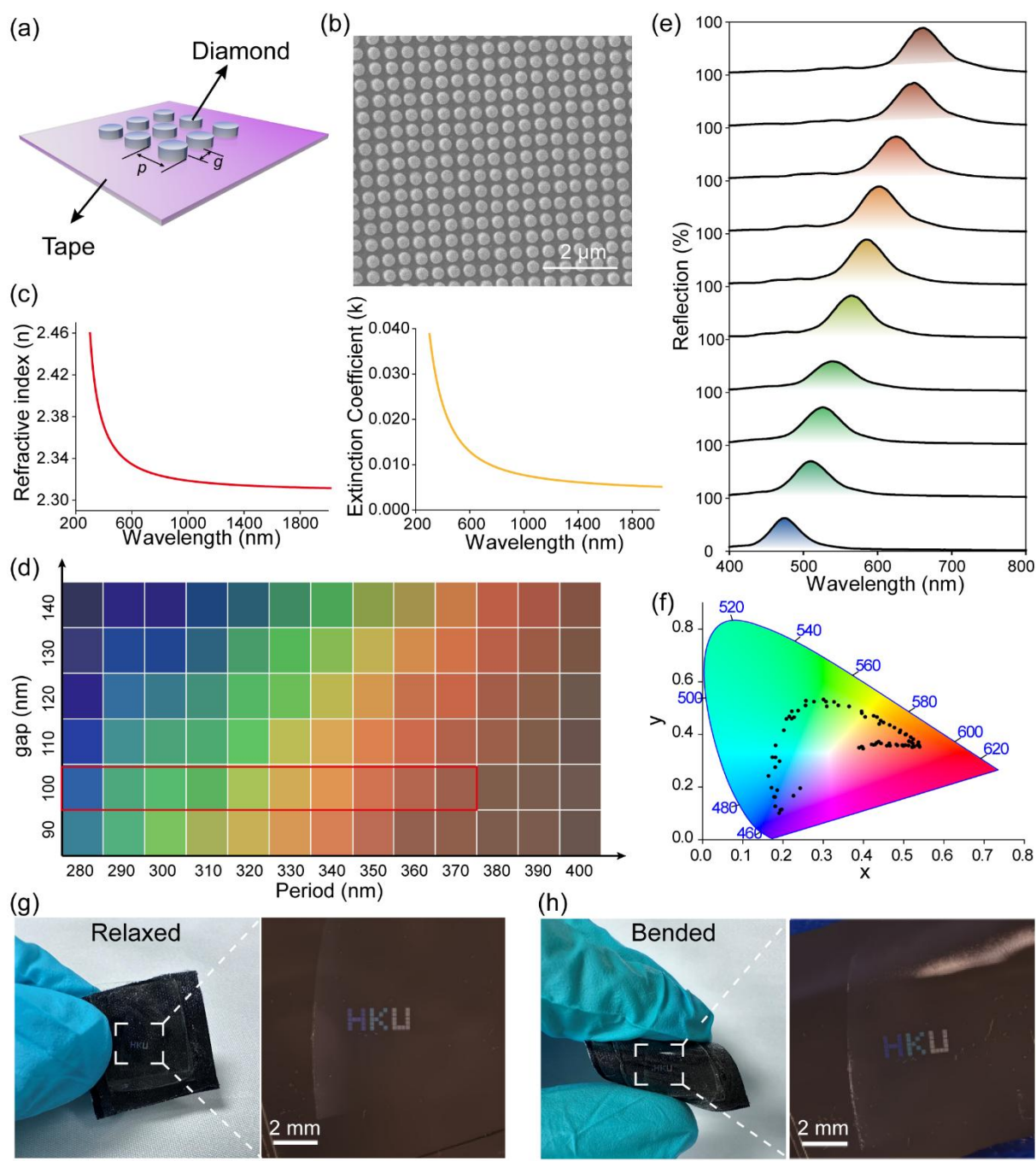

**Figure 4** . (a) Schematic unit cells of diamond metasurfaces on tape substrate. (b) Top-view SEM image of diamond metasurfaces. (c) Measured refractive index and extinction coefficient of the diamond film. (d) Color palette of fabricated diamond structural colors. (e) Corresponding experimental reflectance spectra of diamond structural colors (red mark in (d)), where the period $p$ varies from 280 nm to 370 nm at a fixed $g$ of 100 nm. (f) CIE 1931 color

map of fabricated structural colors. (g, h) Photograph and zoomed in microscope images of diamond structural colors under relaxed and bended status, respectively.

Figure. 4(a) shows the schematic unit cells of diamond metasurfaces, which is composed of diamond pillars and PET tape substrate, where $p$ and $g$ denote period of diamond pillar array and gap between adjacent diamond pillars, respectively. Figure. 4(b) exhibits top-view SEM image of fabricated diamond pillar arrays with the period $p$ = 400 nm and gap $g$ = 100 nm. The measured refractive index and extinction coefficient of the diamond film are shown in Figure. 4(c). Figure. 4(d) presents the color palette of fabricated diamond structural colors by adjusting the period $p$ and gap $g$ from 280 nm to 400 nm and 90 nm to 140 nm, respectively. In particular, the color would change from blue to red as $g$ decreases and $p$ increases, covering the entire visible range. Derived from one row of the color palette (marked in red box), the measured reflectance spectra also behaved the obvious red-shift, as shown in Figure 4(e). Specifically, the reflectance intensity was up to 88.78% when setting $g$ and $p$ as 100 nm and 370 nm. Figure 4(f) exhibits the positions of fabricated diamond structural colors in the standard CIE 1931 color map, which indicates their outstanding color saturation and wide gamut. Due to the flexible PET substrate, the proposed structural color can be bended and conformed to curved surfaces. Figure 4(g, h) present photograph and zoomed in microscope images of diamond structural color under relaxed status and bended status, respectively. Where the diamond structural color is attached onto a piece of black fiber to avoid background light. The subtle difference of reflectance spectra between relaxed and different bending status (Supporting Information Section 4) indicate the proposed diamond structural color possesses superb stability under deformation. The satisfying saturation and ultra-high reflectance intensity compared with previews works,[16,40-43] as well as the stable display characteristics further underscore the superb performance of diamond film and the excellent compatibility of our proposed method for high-precision nano-fabrication, which shows promising potential towards applications including high-performance displays,[43] encryptions,[16,47] flexible and wearable display,[48] durable display devices for harsh environments,[49] etc.

## 4. Conclusion

In this study, we proposed a versatile method for nano-fabrication on our newly developed ultrathin and flexible diamond film, which overcomes the challenges and limitations of conventional nano-fabrication method for two-dimensional materials and bulk materials. Our proposed method involves the straightforward transfer of masks from ITO film to diamond film without the need of extra chemicals, thereby preventing diamond film and substrate from being damaged or contaminated. Compared with masks fabricated by conventional method, our results exhibit significantly higher geometrical resolution and better repeatability, as well as the capability of fabricating large-scale structures. As a demonstration, flexible all-diamond metasurface functioning as structural colors is carried out, which is the first flexible device made with diamond film, the structural colors exhibit remarkable properties including ultra-high reflectance intensity up to 88.87%, wide gamut spanning the entire visible spectrum, satisfying saturation, and stable display under bending. It is believed to hold great promise for stable and high-quality structural colors, wearable displays, and information encryptions. The outstanding performance of the fabricated diamond structural color further substantiate the superiority of the proposed method, unlocking the vast potential applications of diamond films in areas such as diamond photonics, next-generation electronics devices, quantum sensing, heat spreaders, etc.

## 5. Methods

### 5.1. CVD growth of diamond film

The heteroepitaxial growth of diamond on a silicon (Si) substrate encompasses three essential steps: substrate pretreatment, diamond seed deposition, and chemical vapor deposition (CVD) of diamond membranes. Initially, the Si surface underwent hydrogen plasma treatment. A 2-inch silicon wafer was then introduced into a microwave plasma-assisted chemical vapor deposition (MPCVD) device for 10 minutes with settings of 1300-W power, 35-Torr cavity pressure, and a 300-sccm hydrogen ($H_2$) gas flow. Preparing the diamond seeds required a series of mixing, dispersing, and centrifuging steps. Diamond seeds (sourced from Tokyo Chemical Industry Company Limited) smaller than 10 nm were mixed with dimethyl sulfoxide (DMSO), anhydrous ethanol, and acetone at a mass ratio of 1:5000:250:250. This mixture was sonicated for 12 hours to ensure proper dispersion, followed by centrifugation at 1000 rpm for 20 minutes to eliminate impurities. The suspension was then spin-coated onto the Si wafer under the following conditions: an initial spin at 500 rpm with three drops added within 15 seconds, followed by an increase to 4500 rpm for 110 seconds. This spin-coating process was repeated three times. Finally, the wafer with the deposited diamond seeds was placed into the MPCVD setup (Seki 6350) for diamond membrane growth. The primary parameters for growing a 1-μm-thick diamond membrane included 3400-W microwave power, a temperature of 900°C, a 15-sccm methane flow rate, and a 40-minute growth period.

## 5.2. Mask Preparation

A PET film (1.25-mm thick) with 200 nm-surface deposition of electrically-conductive ITO was firstly cut into small pieces of 1 cm × 1 cm size by scissors. The hydrogen silsesquioxane (HSQ) was then spin-coated on the surface of PET film at a spinning rate of 4500 r/min for 60 s, followed by 80 °C heating for 180 s. After spin-coating of E-beam resist, EBL was carried out for pattern definition at a dose of 750 μJ/cm², followed by development in 25% Tetramethylammonium hydroxide (TMAH) for 60 s.

## 5.3. Simulation of diamond metasurfaces

Software Lumerical FDTD solutions was used to calculate the theoretical reflectance of diamond structural color based on finite-difference time-domain (FDTD) method, periodic boundary condition was adopted at the x- and y- direction from top view of diamond structural color to simulate the periodic distribution of diamond unit cells, perfectly matched layer (PML) boundary condition was adopted at the z-axis to serve as absorber to block outgoing electromagnetic wave.

## Competing interest:

The authors declare no competing financial interest.

## Supporting Information

Supporting Information is available from the Wiley Online Library or from the author.

## Acknowledgements

Yicheng Wang and Jixiang Jing contributed equally to this work. Kwai Hei Li acknowledges the Shenzhen Fundamental Research Program (JCYJ20220530113201003). Zhiqin Chu acknowledges the financial support from the National Natural Science Foundation of China (NSFC) and the Research Grants Council (RGC) of the Hong Kong Joint Research Scheme (Project No. N_HKU750/23), HKU seed fund, and the Health@InnoHK program of the Innovation and Technology Commission of the Hong Kong SAR Government.

**Table of contents:**

A versatile, high-precision, non-contaminating sugar-trasnferring method has been developed for nanfabrication on newly developed ultrathin and flexible diamond films. It enables the flexible all-diamond metasurfaces, for the first time, come into the public eyes.

Yicheng Wang[#], Jixiang Jing[#], Yumeng Luo, Linjie Ma, Zhongqiang Wang, Qi Wang*, Kwai Hei Li*, Zhiqin Chu*

**Title: A versatile method for nano-fabrication on diamond film: flexible diamond metasurfaces as a demonstration**

**ToC figure**

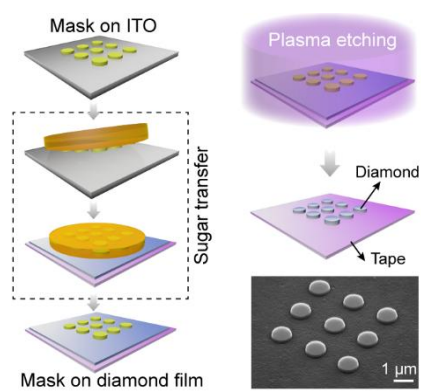

Supporting Information

**A versatile method for nano-fabrication on diamond film: flexible diamond metasurfaces as a demonstration**

*Yicheng Wang[#], Jixiang Jing[#], Yumeng Luo, Linjie Ma, Zhongqiang Wang, Qi Wang\*, Kwei Hei Li\*, Zhiqin Chu\**

**S.1 Method for direct exposure on diamond film's buried surface**

Diamond films were firstly deposited on a 2-inch pre-treated silicon wafer utilizing microwave plasma chemical vapor deposition (MPCVD), thickness ranging from 200 nm to hundreds of microns can be achieved by adjusting the MPCVD recipe. After acquiring diamond film on silicon wafer, Diamond knife is used to create an indent on the back side of silicon wafer, allowing it to be easily broken into desired size, for our structural color demonstration, several millimeters would suffice. Sticky tape is utilized to exfoliate diamond from silicon wafer[1], to facilitate this process, silicon wafer with diamond film is immersed in 20% potassium hydroxide (KOH) solution under 80 degrees Celsius for 10 minutes, this is to dissolve a small amount of silicon at the diamond / silicon interface along the sample's edge. Free-standing diamond film with the size of a few microns at the edge of the sample can then function as a start point for peeling. Figure. S1a. exhibits peeled off diamond film on a highly-transparent tape substrate, where the as-grown surface is attached to tape, thus exposing buried surface for further use. After acquiring diamond-on tape assembly with the size of a few millimeters, it is therefore adhered to a 1 cm x 1 cm silica substrate with another layer of double-sided tape, ensuring the diamond surface to be macroscopically flat. The sample is then spin-coated with 6% Hydrogen silsesquioxane (HSQ) in Methyl isobutyl ketone (MIBK) solution at the speed of 4500 revolutions per minute (rpm) and baked at 80 degrees Celsius for 180 seconds. Ozone treatment for 15 minutes is employed to enhance the hydrophilicity of sample surface, approximately 50 nm of conductive polymer (AR-PC 5092) is spin-coated at the speed of 3000 rpm and baked at 150 degrees Celsius for 120 seconds. 30 keV electron beam lithography (EBL) is used to expose designed exemplary structures including circular arrays with the period of 400 nm, and gratings with the width of 1 micron and gap of 500 nm. Finally, the sample is developed in 25% Tetramethylammonium hydroxide (TMAH) dissolved in deionized water for 60 seconds and rinsed in deionized water for 120 seconds.

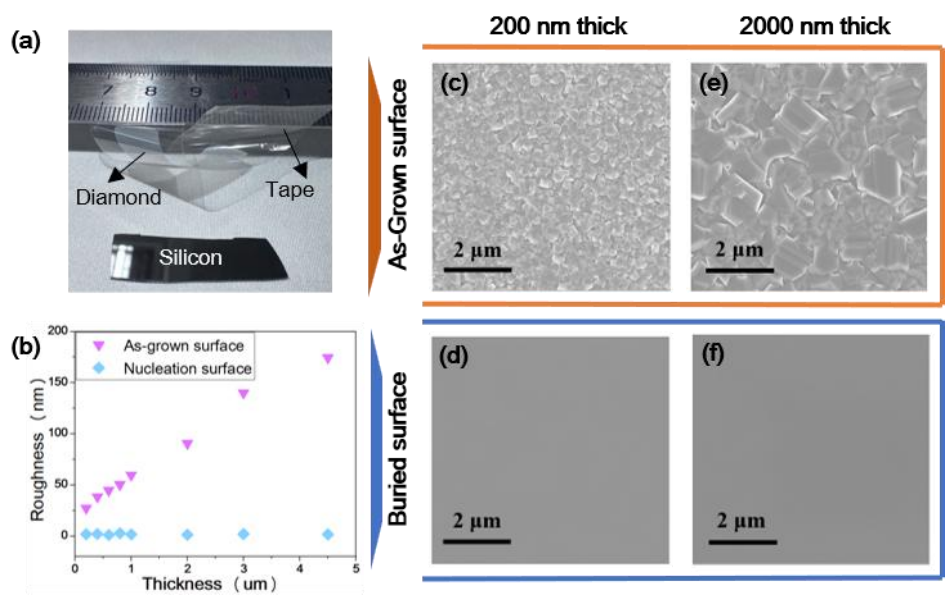

**Figure S1**. (a) Photograph illustrating peeled off diamond film on tape. (b) Measured roughness (Ra) of diamond film's as-grown surface and nucleation surface with thickness ranging from 200 nm to 4.5 microns. (c) and (d) SEM image of 200 nm thick diamond film's as-grown surface and buried surface, respectively. (e) and (f) SEM image of 2000 nm thick diamond film's as-grown surface and buried surface, respectively.

## S.2 Comparison of diamond film's buried and as-grown surface

Atomic force microscope (AFM) was employed to investigate the surface roughness of diamond film. Figure S1b. depicts the relationship between the roughness (Ra) of diamond film's both surfaces and the thickness of diamond film. The roughness of diamond film's buried surface (nucleation surface) is constantly around 1 nm, this value could be further reduced by utilizing finely polished silicon wafer for diamond CVD. The flatness meets the requirement of various applications including fabrication of nano-structures, bonding, etc. Figure S1c. and S1d present SEM image of 200 nm thick diamond film's as-grown surface and buried surface, respectively. And figure S1e, f, present SEM image of 2000 nm thick diamond film's as-grown surface and buried surface, respectively. The high-roughness of diamond film's as-grown surface containing diamond grains with different crystal orientations can introduce instability factors during fabrication procedures, such as spin-coating, bonding, exposure, and etching. In contrast, the ultra-flat property of diamond film's buried surface is significantly more desirable, which is the motivation of detaching diamond film from silicon wafer and exposing the buried surface.

Figure S2 compares fabricated diamond pillars arrays with different geometrical parameters on diamond both surfaces. Diamond films with the thickness of 600 nm were used as an example. Our proposed method was used on both surfaces to transfer masks onto diamond film[2], and inductive coupled plasma (ICP) was employed to etch the film by approximately 200 nm, masks were subsequently removed using hydrofluoric acid. Irregular shaping of diamond pillars and uneven etching were observed on diamond film's as-grown surface, as depicted in figure S2a. and S2b. On the contrary, pillars arrays fabricated on diamond film's buried surface exhibits ultra-flat and uniform properties, making it ideal for the wide range of diamond film's applications.

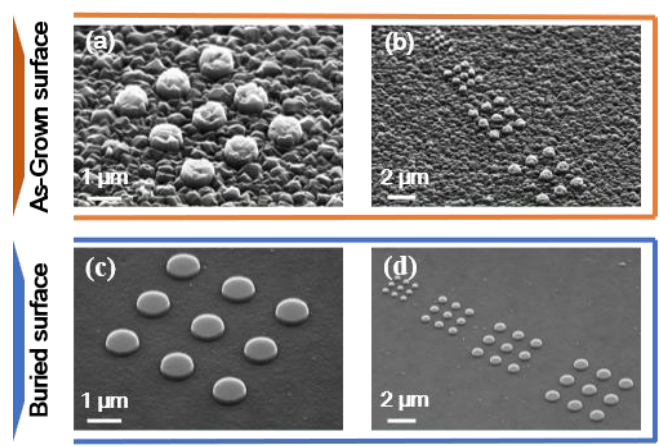

**Figure S2**. (a), (b) SEM images of fabricated pillars array on diamond film's as-grown surface and (c), (d) on diamond film's buried surface, respectively.

### S.3 limitations of using conventional method for nano-fabrications on diamond film's buried surface

### S.3.1. The incompatibility between conventional methods and flexible diamond film

As a newly developed three dimensional (3D) flexible materials, diamond-on-tape assembly need to be adhered onto a rigid substrate during spin coating, photo- / electron beam lithography, and etching, etc. This adherence involves the use and removal of another layer of adhesive or double-sided tape to secure diamond-on-tape assembly to the substrate. This procedure leaves undesired residual of adhesive / double-sided tape and potential chemical contaminations from the removal process. On the other hand, the surface of diamond film on fixed substrate is not strictly flat due to the roughness of tape, adhesive or double-sided tape, wrinkles on diamond film and tapes, and bubbles between each layer. Those factors would significantly affect the uniformity of photoresist (PR) / electron beam resist (EBR) / conductive polymer spin-coated on diamond film. Additionally, directly spin-coating on diamond film intend to create cracks on the film due to the mechanical stress during the procedure, especially for diamond film with the thickness thinner than one micron, which further worsen the film's flatness and integrity. Figure S3a. and S3b presents microscope images of electron beam resist HSQ spin-coated on diamond-on-tape assembly and ITO film, respectively. Where the thickness of diamond is 600 nm, the white box encapsuled a crack on diamond film, and the mottled colors generated by interference of light indicate uneven distribution of spin coated E-beam resist. In contrast, E-beam resist spin-coated on ITO film, as illustrated in Figure S3b. demonstrate excellent uniformity. Moreover, the unevenness of the film poses challenges for focusing laser or electron beam during photo- / electron beam lithography, as well as deviated angle during etching. These problems end up with poor repeatability, distortion of exposed patterns, reduction of resolution, and disability of fabrication large scale structures.

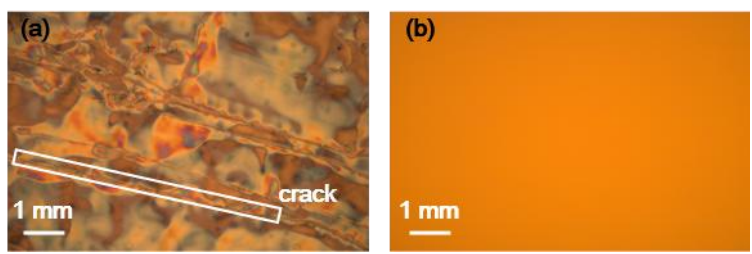

**Figure S3**. Microscope images of HSQ spin-coated on (a) diamond-on-tape assembly and (b) ITO film, respectively.

### S.3.2. The poor conductivity of diamond film

As an insulator, the resistivity of pure diamond without lattice defects is up to $10^{18}$ $\Omega\cdot$m, the value is measured to be $10^{12}$ $\Omega\cdot$m for our diamond film. The high resistivity hinders EBL processes on diamond film due to challenges including charge accumulation and proximity effect. Spin-coating an additional layer of conductive polymer is commonly used to address conductivity issue of samples[3,4], 10 $\Omega\cdot$m to $10^3$ $\Omega\cdot$m can be achieved depending on the types can thickness of conductive polymer. However, as illustrated in previews section, directly spin-coating on diamond-on-tape assembly could end up with uneven distribution of conductive polymer, leading to inconsistent resistivity, and more cracks generated during this process can further undermine the film's integrity, creating open circuits between the two sides of the cracks, which further worsen the conductivity. Additionally, the non-conductive substrate (PET / adhesive) could undergo distortion and deformation during electron beam exposure, introducing more difficulties of sample focusing. Another commonly used method for conductivity issue is depositing a thin layer of metal on sample surface. However, the removal of deposited metal can also involve strong chemicals, which could end up damaging the substrate and contaminating other components of designed devices, yet leaving problems of cracks / uneven surface still unsolved.

**S.4 Stable display under deformation and ultra-high reflectance**

To evaluate the performance of diamond structural colors under different deformations, characterizations in reflectance spectra and color display are investigated, respectively. Figure. S4a presents a schematic diagram illustrating bending flexible diamond structural color sample, where $\theta$ denotes radian of sample area. To demonstrate the stability of our diamond structural color, reflectance spectra were measured under different values of $\theta$. Figure. 4b exhibits normalized reflectance spectra of designed red, green and blue (RGB) colors under the conditions of $\theta = 0°$, $10°$, $20°$, and release ($\theta = 0°$), respectively. Highly stable color performance while bending and releasing with negligible difference of reflectance and was achieved. Figure S4c present a comparison in terms of reflectance intensity between our work and some of structural color works in recent years. Traditional metallic structural colors and dielectric structural colors using silicon and $TiO_2$ were denoted with rectangles, triangles, and circles, respectively. Despite the superior characteristics of diamond, diamond structural colors have been seldomly reported. Our work shows significantly higher reflectance intensity than the previews diamond structural color on silicon substrate, which is denoted in red pentagon. Detailed statistics are shown in Table S1.

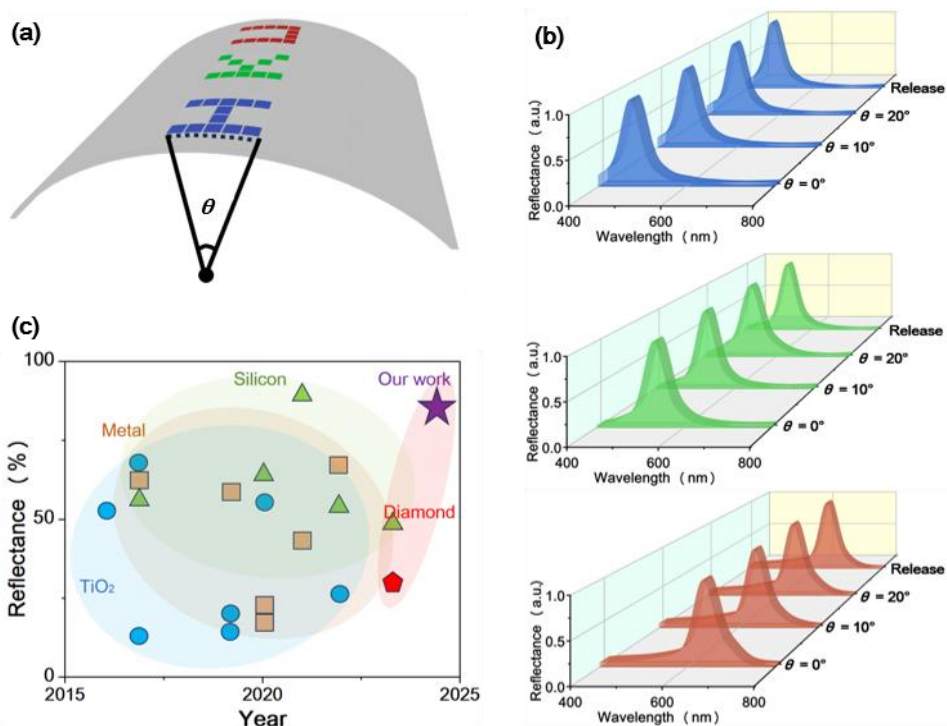

**Figure S4**. (a) Schematic illustration of bending diamond structural color. (b) RGB Reflectance spectra of diamond structural color under the condition of $\theta = 0°$, $10°$, $20°$, and release ($\theta = 0°$), respectively. (c) Comparison of our diamond structural color with structural color utilizing silicon, $TiO_2$, and metal.

**Table S1**.
Statistics of reported works of structural colors in recent years.

| Reflectance (%) | Year | Reference |
|:---:|:---:|:---:|
| 90 | 2021 | S5 |
| 65 | 2020 | S6 |
| 48 | 2023 | S7 |
| 55 | 2022 | S8 |
| 60 | 2017 | S9 |
| 57 | 2020 | S10 |
| 24 | 2022 | S11 |
| 22 | 2019 | S12 |
| 55 | 2016 | S13 |
| 13 | 2017 | S14 |
| 14 | 2019 | S15 |
| 65 | 2017 | S16 |
| 16 | 2020 | S17 |
| 20 | 2020 | S18 |
| 45 | 2021 | S19 |
| 70 | 2022 | S20 |
| 62 | 2019 | S21 |
| 65 | 2016 | S22 |
| 28 | 2023 | S3 |

| 89 | 2024 | This work |